\documentstyle[preprint,aps]{revtex}
\begin{document}
\draft
\title{Hopping Conduction 
in Uniaxially Stressed Si:B
near the Insulator-Metal Transition}
\author{S. Bogdanovich, D. Simonian, S. V. Kravchenko and M. P. Sarachik}
\address{Physics Department, City College of the City University of 
New York, New York, New York 10031}
\author{R. N. Bhatt}
\address{Department of Electrical Engineering, Princeton University, 
Princeton, New Jersey 08544-5263}
\date{\today}
\maketitle
\begin{abstract}
Using uniaxial stress to tune the critical density
near that of the sample, we have studied in detail the low-temperature 
conductivity of p-type Si:B in the insulating phase very near the 
metal-insulator transition.  For all values of temperature and stress, the 
conductivity collapses onto a single universal curve, $\sigma (S, T) = 
AT^{1/2}F[T^*(S)/T]$. For large values of the argument, the scaling function $F[T^*(S)/T]$ 
is well fit by $exp[-(T^*/T)^{1/2}]$, the exponentially activated form associated 
with variable range hopping when electron-electron interactions 
cause a soft Coulomb gap in the density of states at the Fermi energy.  The 
temperature dependence of the prefactor, corresponding to the T-dependence of 
the critical curve, has been determined reliably for this system, and is 
$\propto T^{0.5}$. We show explicitly that neglecting the prefactor leads to 
substantial errors in the determination of the $T^*$'s and the critical 
exponents derived from them.  The conductivity is not consistent with Mott 
variable-range hopping, $exp[-(T^*/T)^{1/4}]$, in the critical region nor does 
it obey this form for any range of the parameters.  Instead, the conductivity 
of Si:B is well fit by $\sigma = AT^{1/2} exp[-(T^*/T)^\alpha]$ 
for smaller argument of the scaling function, with $ \alpha = 0.31$
related to the critical exponents of the system at the metal-insulator transition.

\end{abstract}
\pacs{PACS numbers: 72.20.-i,71.30.+h,72.80.Ng,72.80.Sk}

Hopping conductivity of localized electrons in disordered insulators was a subject of 
considerable controversy two decades ago.  For noninteracting electrons, the problem 
was first addressed by Mott\cite{mottvrh,mottbook}, who showed that below any microscopic 
energy scale, a tradeoff between the exponential thermal activation due to the difference 
in energy between the initial and final electron states on the one hand, and the exponential 
factor associated with the spatial overlap between the two (localized) states on the other, 
leads to an optimal conductivity at low temperatures of the form:

\begin{equation}
\sigma \propto exp [ - (T_0/T)^{1/4} ]
\label{equation}
\end{equation}
in three dimensions, where

\begin{equation}
T_0 \propto  1/[N(E_f) a^3 ]
\label{equation}
\end{equation}
In the above equation, $N(E_f)$ is the (constant) one-electron density of states at the Fermi level, and
$a$ is the (linear) size of the localized electronic wavefunction.
This expression, known as Mott's variable range hopping (VRH),
was put on a rigorous footing using a percolation formalism\cite{ahl,pollak,shklovskiiefros}.
Many different materials appeared to agree well with the Mott
formula\cite{mottexamples}, providing experimental confirmation of Mott's ideas.

The applicability to real disordered insulators was, however, challenged by a number
of theorists\cite{pollak2,srini}, because of the presence of Coulomb
interactions between electrons. A key step in understanding the
role of electron interactions was put forward by Efros and Shklovskii (ES)\cite{efrosshklovskii}, who 
showed that a
self consistent Hartree treatment of the long range (1/r) Coulomb interactions in
an insulator leads to a soft gap in the one-electron density of states at the Fermi level,
resulting in a depletion of low lying excitations.  This, in turn, leads to a much lower 
conductivity at low temperatures of the form:

\begin{equation}
\sigma \propto exp [ - (T'_0/T)^{1/2} ]
\label{equation}
\end{equation}
where

\begin{equation}
T_0' \propto e^2/{\epsilon a}
\label{equation}
\end{equation}
Here $e$ is the electronic charge, $\epsilon$ the dielectric constant of the semiconductor, and $a$ the linear size
of the localized electronic state.

Considerable activity on the issue ensued in the years following, during which various materials were 
shown to obey either the Mott form (Eq. 1) or the ES form (Eq. 3).  A crossover with decreasing 
temperature from Mott to ES variable-range hopping was observed in CdSe:In\cite{zhang} and 
CdTe:Cl\cite{agrin}.  This was 
attributed\cite{zhang,aharony} to hopping energies that were larger than the gap energy at high temperature (Mott 
hopping) and smaller than the gap at low T (ES hopping).  A crossover with dopant 
concentration was found in n-GaAs\cite{tremblay}, where Mott hopping was claimed for samples near the metal-insulator 
transition when the Coulomb gap has a small energy-width, and ES hopping prevails deeper in 
the insulating phase where electron-electron interactions are stronger and the hopping 
electrons probe the gap.  Although variable-range hopping exponents have been found that deviate from 
these values, it is found that strong electron interactions yield a hopping 
exponent of 1/2 
while weak interactions (compared with hopping energies) give rise to exponent 1/4. 
This has given rise to the expectation that Mott variable-range hopping will always be observed 
near the metal-insulator transition as electron screening increases and the Coulomb gap 
collapses approaching the metallic phase\cite{tremblay,shafarman}.

In this paper we report measurements of the hopping conduction in insulating Si:B very near
the transition to the metallic phase. By applying a compressive uniaxial stress along the
[001] direction using a pressure cell described elsewhere\cite{bogdanovich}, we have driven a sample of
Si:B from the metallic phase into the insulating phase, and mapped out the conductivity
as a function of applied stress (S) and temperature (T) in the range $0.05 K < T < 0.5 K$.
We find\cite{ourPRL} that the conductivity over this entire temperature range for stress values
varying by  about $40$ percent on either side of the critical stress $S_c$, is described accurately by the 
scaling form:

\begin{equation}
\sigma (S, T) = \sigma_c(T) f[\Delta S ~ T^{-1/z\nu}],
\label{eq}
\end{equation}
where $\sigma_c(T)$ is the conductivity 
at the transition, $\Delta S = (S_c - S)$ is the difference between the stress and 
its critical value (i.e the control parameter),  $\nu$ is the critical exponent 
that characterizes the divergence of the correlation length, 
$\xi \propto (\Delta S)^{- \nu}$, and $z$ is the dynamical exponent that describes the divergence of 
the time scale, $\tau \propto \xi^z$.  By defining a stress-dependent temperature scale 
$T^* \propto ( \Delta S )^{z\nu} $, and noting
from our previous work\cite{ourPRL} that $ \sigma_c (T) \propto T^{1/2} $, we may rewrite the above
equation as:
\begin{equation}
\sigma (S,T) = A T^{1/2} F[T^*/T]
\label{equation}
\end{equation}
This equation fully
describes the conductivity of Si:B on both the insulating and metallic sides in the  
vicinity of the metal-insulator transition. We present below a detailed analysis of  the temperature dependence of
the conductivity of the insulating branch in the critical region near the transition.

	A bar-shaped $8.0$x$1.25$x$0.3$ mm$^3$ sample of Si:B was cut with its long 
dimension along the [001] direction.  The dopant concentration, determined 
from the ratio of the resistivities\cite{bogdanovich} at 300 K and 4.2 K, 
was $4.84$x$10^{18}$ cm$^{-3}$.  
Electrical contact was made along four thin boron-implanted strips.  Uniaxial 
compression was applied 
to the sample along the long [001] direction using a pressure cell described 
elsewhere\cite{bogdanovich}.  Four-terminal measurements were taken at 13 Hz (equivalent 
to DC) for different fixed values of uniaxial stress at temperatures between 
0.05 and 0.76 K.  Resistivities were determined from the linear 
region of the I-V curves.

For two Si:B samples with different dopant concentrations that are  metallic 
in the absence of stress, Fig. 1 shows the resistivity at 4.2 K normalized to its 
zero-stress value as a function of uniaxial stress.  With increasing stress, 
the resistivity initially 
increases rapidly (by several orders of magnitude) and 
then decreases gradually above several kilobar.  This 
is in marked contrast with Si:P, which exhibits little change at small stress 
values, and then shows a similar decrease in resistivity at larger stresses.  
This can be understood as follows.  
The acceptor state in Si:B has a four-fold degeneracy in the unstressed cubic 
phase which is lifted by relatively small uniaxial stress into two doublets
(each retaining only the Kramers degeneracy); this initially drives Si:B more 
insulating.  By contrast, the six-fold valley degeneracy (on top of the 
required Kramers or spin degeneracy) of an effective mass donor in Si has 
already been removed (even in zero stress) by the central-cell potential of 
the phosphorus dopants\cite{bhatt}. Such contrasting behavior is due in part
to wavefunction anisotropy\cite{bhatt2} and in part to degeneracy in the presence of
electron correlation\cite{bhattandsachdev}, whose effects have been separately
considered for the case of effective mass donor systems.

	The conductivity is shown as a function of temperature on a log-log 
scale in Fig. 2 for different uniaxial stresses for which the sample is in the 
insulating phase.  Based on a detailed analysis published elsewhere\cite{ourPRL}, 
the critical stress for this sample was determined to be $S_c = 613$ bar.  
The critical curve is a straight line on this scale, with 
the conductivity $\sigma_c \rightarrow 0$ at $T \rightarrow 0$, 
following a power law, $\sigma \propto T^{0.5}$.

The conductivity $\sigma (S, T)$ normalized 
by the critical conductivity $\sigma_c (T)$ is shown in Fig. 3 as a function of 
the scaling variable,
$\Delta S/T^{1/z\nu}$ where $z\nu=3.2$ has been chosen so that the data 
for all values of stress and all measured temperatures collapse onto a single 
universal curve, as predicted by Eq. (1).  The resulting scaling function fully 
describes the temperature dependence of the conductivity in the insulating phase
in the vicinity of the transition.
 
As discussed earlier,
the conductivity in the insulating phase is expected to exhibit variable-range
hopping at low temperature of the form:
\begin{equation}
\sigma (S, T) \propto \sigma_0 (T) exp[-(T^*/T)^\alpha] ,
\label{eq}
\end{equation}
with $\alpha = 1/4 $ when the density of states is a slowly varying function of energy 
(Mott-VRH\cite{mottvrh,mottbook}, and $ \alpha = 1/2 $  (ES-VRH\cite{efrosshklovskii})
when hopping energies are comparable with or smaller 
than electron interactions, forming a soft ``Coulomb" gap at the Fermi level.  
While these analyses have been done for the strongly localized regime (deep in the
insulating phase), arguments have been advanced\cite{bhatt3}
why such behavior persists in the insulating phase even close to the transition,
provided the temperature is low enough that the localized character of the phase
becomes evident.

However, it is not clear whether the hopping conduction is included in the scaling part of
the conductivity near the metal-insulator transition. It has been suggested\cite{polyakov}
that it is in the case for of quantum Hall transition seen in two-dimensional electron
gases in the presence of a strong perpendicular magnetic field, but the experimental
evidence for this is not unambiguous. For the metal-insulator transition in
three dimensions it is clear from Eq. (4) that in order that VRH be part of the scaling
description, we must have
$\sigma_0 (T) = \sigma_c \propto T^{1/2}$, $f[T^*/T] \propto exp[-(T_0/T)^\alpha] $ and 
$T^*\propto T_0\propto (\Delta S)^{z\nu}$.

To test if the scaling description contains either of
these two forms of hopping conduction, we plot the conductivity normalized 
to the conductivity of the critical curve, $\sigma/\sigma_c$, on a log scale as a function  of 
$(T^*/T)^{1/2}$ in Fig. 4a and as a function of $(T^*/T)^{1/4}$ in Fig. 4b.  
As can be clearly seen in Fig. 4a, the experimental data
for $ (T^*/T)^{1/2}>2.8 $ ( $T^*/T>8 ) $ lie on a straight line passing through the origin, indicating
that the conductivity crosses over to an ES-VRH form within the scaling region at large
but experimentally accessible values of the argument of the scaling function,
with a temperature-dependent prefactor given 
by the critical curve, namely:
\begin{equation}
\sigma (T) \propto T^{1/2} exp [-(T^*/T)^{1/2}].
\label{equation}
\end{equation}

Deviations are evident for $T^*/T \leq 8$.  
For such small arguments of the exponential factor, it has been argued for the insulator
that hopping energies may be comparable or larger than the 
energy-width of the Coulomb gap; in this regime a crossover has been observed 
in some systems, albeit within limited range\cite{zhang,agrin} to Mott VRH with an exponent 
$1/4$ rather than $1/2$.  However, it is clear from the consistently downward curving plot in Fig 4b that Mott hopping
does not provide an adequate description of the conductivity of uncompensated 
Si:B in the critical region for any range of $T^*/T$,
particularly if one realizes that the curve must pass through the upper left corner.

What, then, is the form of the scaling function for $ T^*/T \leq 8 $ ?
In a scaling description of a continuous phase transition, the singular behavior of
the system in the vicinity of the transition is embodied in non-trivial
but universal exponents, as well as ratios and combinations of variables which have non-analytic
form at the approach to the transition, but with scaling functions that are themselves analytic
functions of these ratios or combinations. Consequently, we would expect the scaling function $f$
in Eq. (5) to be analytic in its argument around the origin. Given that $f(0) = 1$, and that we 
expect it to decrease 
monotonically to its asymptotic value
$ f (\infty) = 0 $ on the insulating side, a reasonable choice for $y \geq 0$ is  $ f(y) = exp (- \gamma y ) $, 
suggesting that
\begin{equation}
\sigma (T) \propto T^{1/2} exp [-(T^*/T)^{1/z\nu}].
\label{equation}
\end{equation}
{\em i.e.}, the normalized conductivity $ \sigma / \sigma_c $ should yield a straight line when 
plotted on a semilogarithmic scale versus
$ (T^*/T)^{1/z\nu} $. That this is the case is shown in Fig 5;
however, since the conductivity varies by only one order of magnitude in this range, 
it might be argued that the function could perhaps be equally well described by a power-law form
consistent with the boundary conditions $f(0) = 1$ and $f(\infty)= 0$, {\em e.g.},
\begin{equation}
\sigma(T) / \sigma_c (T) = [1 + (T^*/T) ^ \beta] ^ {-1} 
\label{equation}
\end{equation}
To test whether such a fit works well, we have plotted in the inset to Fig 5 the expression
$ [(\sigma_c / \sigma) - 1] $ versus $ T^*/T $ on a double logarithmic plot, where a power 
law should yield a straight line. This provides a reasonable fit over 
a much smaller range (note the logarithmic scale), suggesting that the 
scaling function is better described by an 
exponential than any single power-law; indeed, it fits
over much of the range of the argument of the scaling function before it crosses over
to ES-VRH (Fig 4a).
By combining data for the temperature-dependent conductivity for a number of
values of uniaxial stress, we have thus been able to map the scaling function for a
large range of its argument. We have established that for Si:B, the
conductivity in the insulating phase in the scaling region appears to be equal to a 
prefactor given by the power law behavior of the critical conductivity, multiplied by an 
exponential function of
$(T^*/T)$ raised to a power $\alpha$
which equals $1/z\nu$ for small argument, and $1/2$ for large value of the argument, 
corresponding to ES-VRH.

In analyzing the conductivity of the insulator in the VRH regime,
the temperature-dependent prefactor is very often omitted 
because its weak (power-law) dependence is negligible compared 
to the strong temperature-dependence of the exponential term.  This is certainly 
justified deep in the insulator; however, its neglect is questionable
in the critical regime, where fits to the ES-VRH form have been used to extract
critical exponents pertaining to the insulator-metal transition. We now show
explicitly that omission of this term near the transition in our data leads to 
significant errors in the determination of $T^*=T_0'$ and the critical 
exponents derived from them.

Applying the usual analysis for ES hopping which neglects the 
temperature-dependent prefactor, we plot $\sigma$ on 
a log scale versus $T^{-1/2}$ in Fig. 6a.  A reasonably good fit (i. e., 
straight line) is obtained for the higher values of stress; not unexpectedly, 
deviations become progressively more pronounced as the transition is approached.  
Except very near the transition, the conductivity appears to be well-described 
by ES-VRH, for which
\begin{equation}
T_0' \propto 1/(\epsilon \xi) \propto (\Delta S)^\alpha.
\label{eq}
\end{equation}
Here $\xi$ is the correlation length and $\epsilon$ is the dielectric constant.  
The inset to Fig. 6a shows a plot of $T_0'$ derived from this analysis versus 
$S$, yielding $\alpha = 2.8$.  Since $\alpha$ plays the same role as 
$z\nu$ which was found to equal $3.2$ in the earlier analysis, neglect of the 
temperature-dependence of the prefactor gives rise to an error on the order of 
15 percent in the determination of the critical exponents.  For comparison and 
completeness, we plot in Fig. 6b the correct form, $\sigma/T^{1/2}$ versus 
$T^{-1/2}$.  Inclusion of the prefactor,as in Fig. 6 (b), provides a much better fit 
over a wider range to ES-VRH than does the neglect of the prefactor, as in Fig 6 (a).  
Moreover, it yields a different value for the critical exponent $ 1/z\nu $.
We caution, however, that comparison
with the full scaling curve of Fig. 3 reveals that (smaller) deviations occur 
here due to departures from the ES hopping form as the transition is approached.  We suggest that 
a reliable determination of the prefactor requires a full scaling analysis of 
the conductivity, and cannot be obtained from the individual curves.

	To summarize, we have shown that scaling provides an excellent description of 
the conductivity near the metal-insulator 
transition in uniaxially stressed Si:B.  Based on data at many values of 
stress and temperature, the scaling function in the insulating phase yields 
particularly reliable determinations of the conductivity 
in the critical region. It is found that the conductivity
expected for variable range hopping in the presence of Coulomb interactions,
in the form predicted by Efros and Shklovskii, is part of the scaling description
in the insulating phase for large values of the scaling argument ({\em i.e.} temperatures
T an order of magnitude lower than the characteristic temperature $T^*$). For lower
values, a clear deviation is seen, and seems to be well fit by an exponentially activated
form with an exponent $ 1/z\nu $, which is found to be 0.31 for Si:B.
It would be of interest to see if similar behavior is found in other systems near the
metal-insulator transition, and to check if some of the earlier crossovers seen from ES to Mott VRH
could be manifestations of the same effect.
We have also examined the errors
associated with analysis of conductivity data based on individual curves and 
neglecting the temperature dependence of the prefactor.

We thank L. Walkowicz for her valuable experimental contributions.  We are 
grateful to G. A. Thomas for his generous support and expert advice, help and 
interest throughout this project.  We thank T. A. Rosenbaum, M. 
Paalanen, E. Smith and S. Han for valuable experimental tips and information 
and the loan of equipment, F. Pollak for useful suggestions 
and some samples, and D. A. Huse for a discussion on scaling functions.  This work was supported by the 
US Department of Energy Grant No.~DE-FG02-84ER45153.  R. N. B. was 
supported by NSF grant No. DMR-9400362, and thanks the Aspen
Center for Physics for hospitality while this paper was being written.

\begin{figure}
\caption{For two Si:B samples with different dopant concentrations, as labelled, 
the resistivity at 4.2 K normalized to its zero-stress value as a function 
of uniaxial stress along the directions indicated.}
\label{fig1}
\end{figure}
\begin{figure}
\caption{The conductivity as a function of temperature on a double logarithmic 
scale for various values of uniaxial stress, as labelled, that place the sample 
on the insulating side of the metal-insulator transition.  The critical stress, 
$S_c$, is 613 bar.}
\label{fig2}
\end{figure}
\begin{figure}
\caption{On a log-log scale, the normalized conductivity, $\sigma/\sigma_c$, as 
a function of the scaling variable $[(\Delta S)/S_c]/T^{1/z\nu}$, with $z\nu = 
3.2$.  Here $\Delta S = (S - S_c)$, where $S_c$ is the critical stress.}
\label{fig3}
\end{figure}
\begin{figure}
\caption{(a)  The normalized conductivity, $\sigma/\sigma_c$, on a logarithmic scale 
as a function of $(T^*/T)^{1/2}$; here $T^*\propto (\Delta S)^{z\nu}$; (b) 
The normalized conductivity, $\sigma/\sigma_c$, on a log scale 
as a function of $(T^*/T)^{1/4}.$}
\label{fig4}
\end{figure}
\begin{figure}
\caption{(a)  The normalized conductivity, $\sigma/\sigma_c$ on a logarithmic 
scale as a function of $(T^*/T)^{1/z\nu}$ with $z\nu = 3.2$.  To illustrate 
that a power law does not provide an equally good fit, the inset shows 
$[(\sigma_c/\sigma) - 1]$ as a function of $(T^*/T)$ on a log-log scale 
(see text).}
\label{fig5}
\end{figure}
\begin{figure}
\caption{(a)  Conductivity  $\sigma$ on a log scale as a function of $T^{-1/2}$; the 
inset shows $T_0'$ derived from the slopes in the main figure plotted as a function of 
stress.  (b) The normalized conductivity, $\sigma/T^{-1/2}$, on a log scale as a function 
of $T^{-1/2}$.}
\label{fig6}
\end{figure}

\end {document}
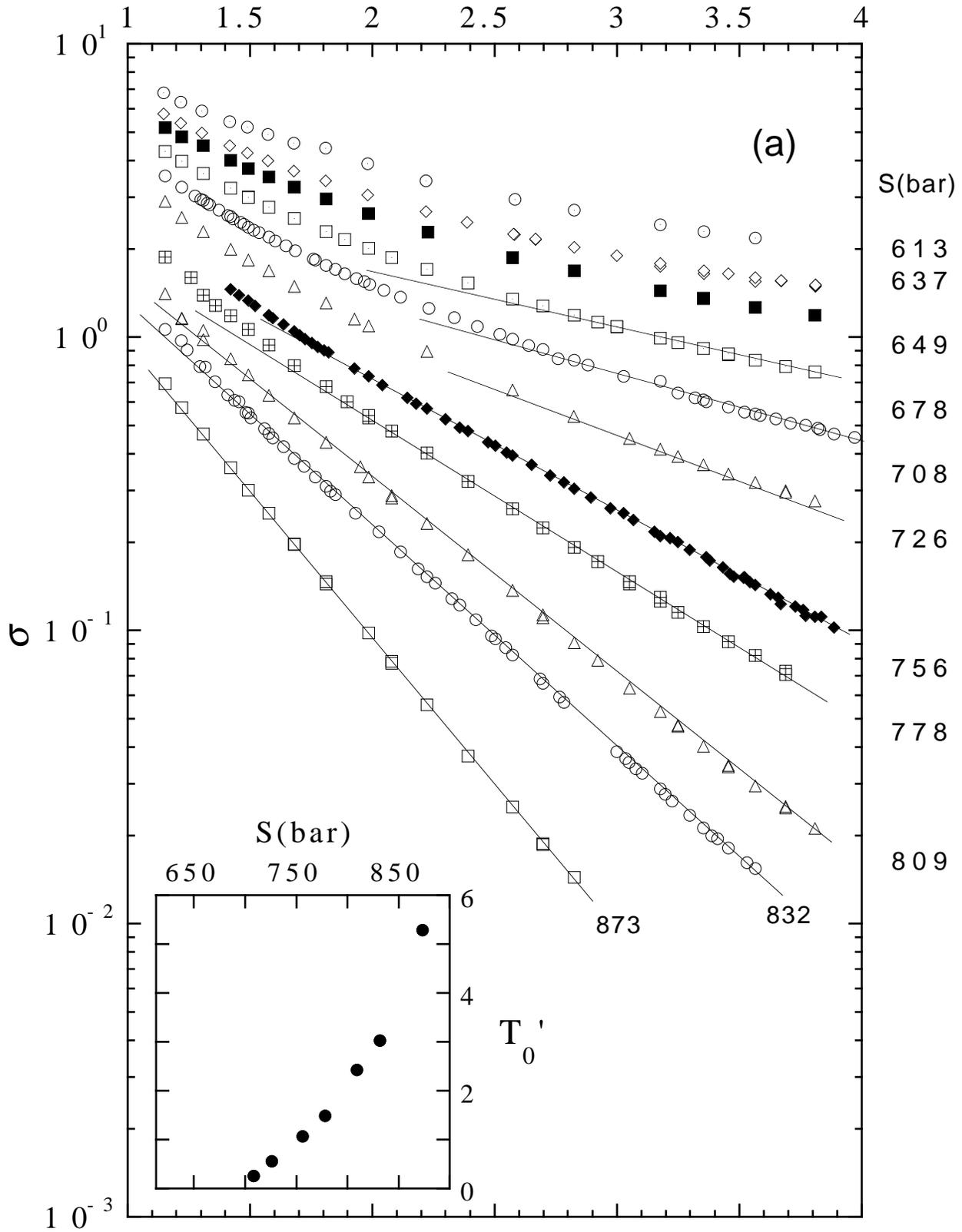
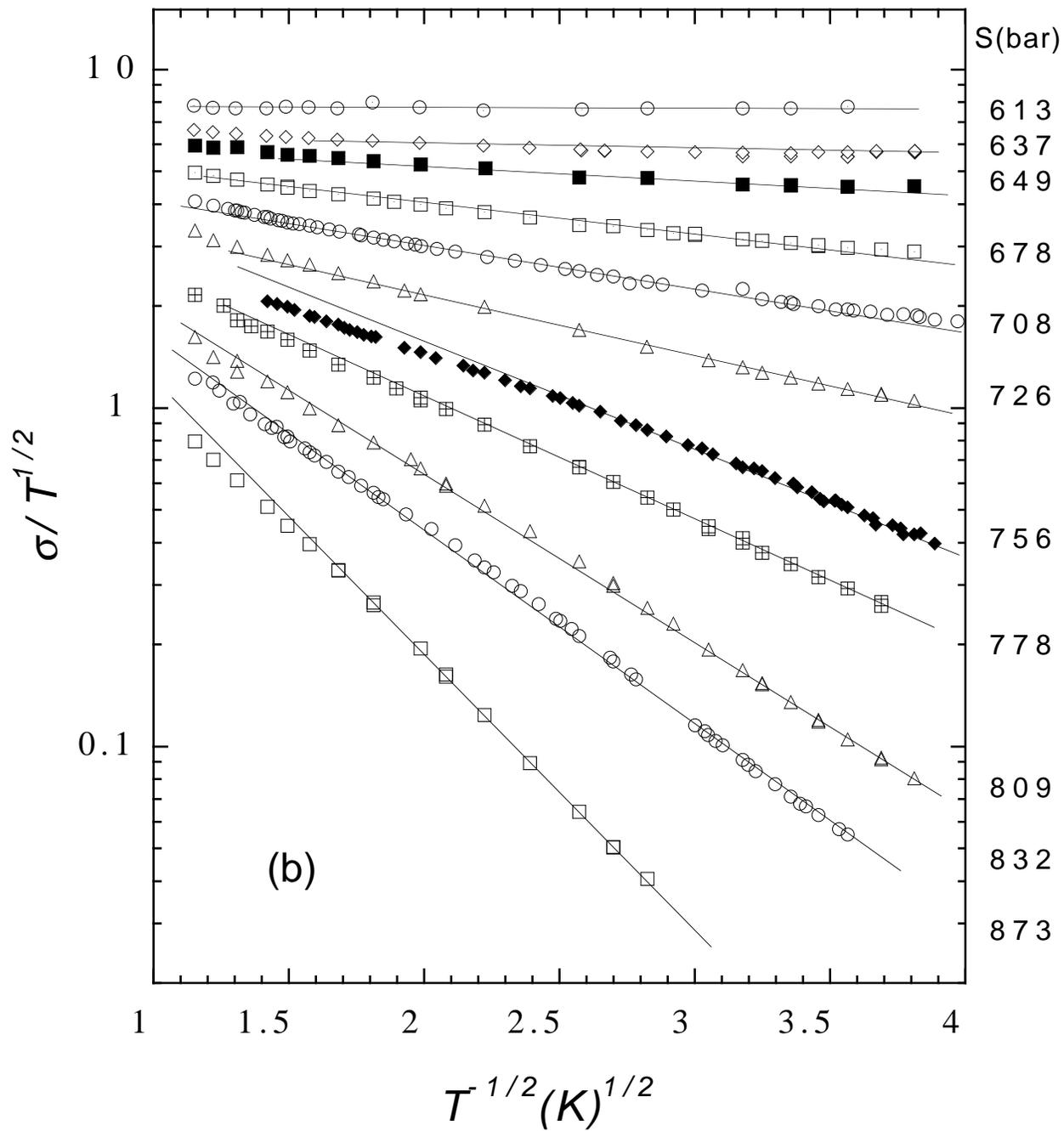
